# Few pulses femtosecond laser exposure for high efficiency 3D glass micromachining

ENRICO CASAMENTI*, SACHA POLLONGHINI AND YVES BELLOUARD

*Galatea Laboratory, IEM/STI, Ecole Polytechnique Fédérale de Lausanne (EPFL), Rue de la Maladière 71b, CH - 2000 Neuchâtel, Switzerland*
*\* enrico.casamenti@epfl.ch*

**Abstract:** Advanced three-dimensional manufacturing techniques are triggering new paradigms in the way we design and produce sophisticated parts on demand. Yet, to fully unravel its potential, a few limitations have to be overcome, one of them being the realization of high-aspect-ratio structures of arbitrary shapes at sufficiently high resolution and scalability. Among the most promising advanced manufacturing methods that emerged recently is the use of optical non-linear absorption effects, and in particular, its implementation in 3D printing of glass based on femtosecond laser exposure combined with chemical etching. Here, we optimize both laser and chemical processes to achieve unprecedented aspect ratio levels. We further show how the formation of pre-cursor laser-induced defects in the glass matrix plays a key role in etching selectivity. In particular, we demonstrate that there is an optimal energy dose, an order of magnitude smaller than the currently used ones, yielding to higher process efficiency and lower processing time. This research, in addition to a conspicuous technological advancement, unravels key mechanisms in laser-matter interactions essential in chemically-based glass manufacturing and offers an environmentally-friendly pathway through the use of less-dangerous etchants, replacing the commonly used hydrofluoric acid.

## 1. Introduction

Laser-based method for fabricating three-dimensional structures at micro/nano-scales has been a long time endeavor embraced by researchers since the 80s. From advanced stereolithographic concepts [1,2] to the reporting of sub-laser wavelength resolution in polymers [3,4], it rapidly expanded towards glass materials at the turn of the millennium, with the demonstration of micron to sub-micron femtosecond laser processing of glass [5–8]. This two-step process consists of first, exposing a substrate to femtosecond laser irradiation to define patterns of arbitrarily shapes throughout the material volume, and second, placing the substrate in an etchant that dissolves laser-exposed volumes. The laser does not remove any material but instead locally modifies it, introducing self-organized nanostructures [9] consisting in a series of nano-planes parallel one to another that are preferentially etched according to their orientation [10]. Contrary to laser induced photo-polymerization that defines the final shape of the object by direct-laser writing, the glass-based exposure-etching method is a negative process as the laser-exposed volume is removed, and what is left is the non-exposed region. As a result, while sub-wavelength resolution exposure patterns can be imprinted using just-above-the-threshold pulse energies, the ultimate feature size and process capability depend not only on the chemical etching selectivity between laser-affected zones and non-affected ones, but also on the etchant diffusion process since its concentration is depleted as it infiltrates the material through the laser-exposed volumes. Achieving the highest possible aspect ratio with the smallest possible feature sizes is therefore directly correlated to the ability to produce laser-modified zones exhibiting the highest possible etching rate enhancement with respect to the etching rate of the pristine material, and yet, with the lowest possible pulse energies so that the non-linear affected zone remains the smallest possible.

To date, two chemical solutions have been reportedly used. The first one consists of using a low-concentration (typically 2.5 to 5 v%), room-temperature-bath of hydrofluoric acid (HF) [6], while the second is to use a high molar concentration solution (8 to 10 M) of potassium hydroxide (KOH) brought at a temperature between 80 and 90 °C [11]. While HF achieves etching contrast in the order of 1:50 to 1:100, KOH exhibits significantly higher values [12]. Both methods pose difficulties in terms of implementation, not only for the operators' safety but also in terms of environmental impact. HF acid is one of the most hazardous acids to manipulate and requires specific recycling procedures. KOH, as all alkali metal hydroxides, remains highly corrosive and hazardous when used at high concentrations.

Here, we investigate sodium hydroxide (NaOH) as an etching solution. Our observations demonstrate an etching contrast, four and two times higher than HF and KOH, respectively. The highest etching rate with NaOH is also observed for very low net exposure doses (~ 1.5 J/mm$^2$), an order of magnitude lower than the ones conventionally used. This specific observation enables ten-fold accelerated laser-exposure velocity than the current state of the art [13] as fewer overlapping pulses are required. We further demonstrate that at least in the very low exposure dose regime, etching enhancement mechanisms by HF, NaOH, and KOH are driven by the presence of defects such as non-bridging oxygen hole (NBOHC) and oxygen deficiency centers (ODC), and not primarily by the presence of nanogratings [14].

## 2. Experiments

*2.1 Protocol*

Fused silica substrates (Corning 7980 0F, OH 1000 ppm, 1 mm-thick, and 25 mm-square) are exposed to a femtosecond laser in a regime where no ablation occurs. In practice, we use an Ytterbium-fiber amplifier laser (Yuzu from Amplitude), emitting 270 fs-pulses at a wavelength of 1030 nm and a constant repetition rate of 333 kHz, chosen far-away from the regime where thermal accumulation is observed (~ 1 MHz). Laser-exposure patterns are written by translating the substrate with linear stages (PI-Micos, UPS 150) under a 0.4-numerical aperture microscope objective that focuses the beam down to a measured optical waist of 1.94 µm. The etching rate efficiency is assessed by measuring the etchant progression in straight line patterns passing through the entire specimen and buried under the material surface at a fixed depth of 70 microns. A digital microscope (Hirox KH-8700) is used to measure the length of the etched patterns after an etching time of four hours for the three etchants considered here.

Lines-patterns are written in back and forth directions, and under two different linear polarization states, aligned and perpendicular to the writing direction, respectively. The pattern lengths are chosen to exceed the actual specimen size to exclude acceleration and deceleration phases of the moving stages, ensuring a constant cruising speed and hence, a constant exposure dose throughout the specimen.

Although near the absorption threshold (found at pulse energies of ~ 160 nJ), the laser-pattern width at the focal point can effectively be smaller than the spot size itself (due to non-linear absorption effects), in what follows, for comparative purpose, we use the optical beam waist as the metric for calculating the net laser exposure dose (or deposited energy) according to the following formula [13]:

$$\Phi = \frac{4 E_p f}{\pi w v}$$

Where $\Phi$ is the exposure dose (in J/mm$^2$), $f$ the repetition rate of the laser (in Hz), $w$ is the optical beam waist (in mm), $v$ is the velocity of translation of the substrate (in mm/s) and Ep the pulse energy (in J).

A quantitative comparison between hydrofluoric acid (HF, 2.5 v%), potassium hydroxide (KOH, 45 wt%), and sodium hydroxide (NaOH, 5 wt%) as chemical etchants is performed. HF etching is carried at room temperature, while KOH and NaOH etching are performed at 90 °C. Six different pulse energies are explored from 160 to 260 nJ (with a step of 20 nJ), spanning across a region where the laser modified zone stretches from ~ 6 µm to ~ 17 µm along the laser propagation axis. These pulse energy conditions correspond to a pulse-irradiance $\psi = E_p / (\Delta t \cdot \pi w^2 / 4)$, where $\Delta t$ is the pulse duration (in s), of ~ 200 to ~ 326 W/mm$^2$.

For each pulse energy, the exposure dose is varied 30 times between ~ 0.5 and ~ 100 J/mm$^2$ by tuning the laser-scanning speed between 0.5 and 85 mm/s, while keeping the repetition rate constant (333 kHz). The samples are cut into two perpendicularly to the sets of lines to expose the laser affected zones directly to the etchant, avoiding the beam-clipping effects that occur on the edges of the specimens. In this fashion, two samples are produced for each specific combination of parameters for statistical purposes.Comparison between etching solutions: HF, KOH and NaOH

*2.2 Effect of pulse energy*

Fig. 1 shows the etching rate comparison of the three etchants for a given set of pulse energies and as a function of the exposure dose for a fixed polarization, chosen perpendicular to the writing direction. This polarization is reportedly the one that leads to the highest etching rates [10].

The three etchants follow a similar overall trend, with noticeable and significant differences in the moderate pulse energy regime, found around 240 nJ. For the near-absorption threshold pulse energy (~160 nJ), a single peak of etching efficiency is observed at doses around 10 J/mm$^2$, followed by a decay. There, the three etchants display a remarkably identical etching efficiency, with NaOH decaying at a faster rate towards higher doses. This trend - a sharp increase followed by a decay - is consistent with what we reported in [13].

As the pulse energy is further increased, we observe the gradual appearance of a second peak at significantly lower doses, centered on 1.5 J/mm$^2$. While this peak eventually reaches the same amplitude that the one at 10 J/mm$^2$ for HF, it is double in magnitude for the bases, KOH and NaOH. In general, NaOH and KOH etching efficiencies are higher than HF for nearly all the exposure doses, with NaOH reaching the highest recorded etching rate value of >300 µm/h, about four times more than for HF and twice more than KOH under similar exposure conditions. Fig. S1 in the *Supplementary material* shows the sets of lines written with 260 nJ of pulse energy after etching.

A salient feature, common to all three etchants, is the presence of a valley, where the etching rate efficiency significantly drops before recovering to higher values. This region is consistently found around 3-4 J/mm2 and defines a transition zone, signaling a dramatic change in the mechanism promoting an accelerated etching in laser-affected zones between low (1.5 J/mm2) and high (10 J/mm2) exposure doses. It is remarkable to observe that the highest etching rate is observed at the lowest doses, where only a few pulses overlap, in a number not sufficient for the formation of clear distinguishable nanogratings to be observed [15,16].

Fig. 2A shows the lines aspect ratio, i.e. the ratio between the measured etched length and the width at the entrance point of the etched tunnel (see inset of Fig. 2C), for the two characteristics etching rate peaks (~ 1.5 J/mm2 and 10 J/mm2, respectively) defining local maxima, for the three etchants and for each pulse energy. While HF seems to show a weak dependence on the pulse energy, both KOH and NaOH show a clear maximum around 220-240 nJ. This maximum is particularly pronounced for NaOH - where an impressive aspect ratio approaching 400 is observed, while it does not exceed 100 and 50 for KOH and HF respectively (consistently with what previously reported in [11,13]). To visualize the difference between the etchants, grid structures are laser-inscribed and etched for around 18 hours and shown in Fig. 2B. Here the access for the etchant is defined by a vertical plane that intersects the lines. It is clear that NaOH and KOH etch the laser-modified lines faster and are more affected by the laser polarization than HF and, at the same time, NaOH is the most selective of the three.

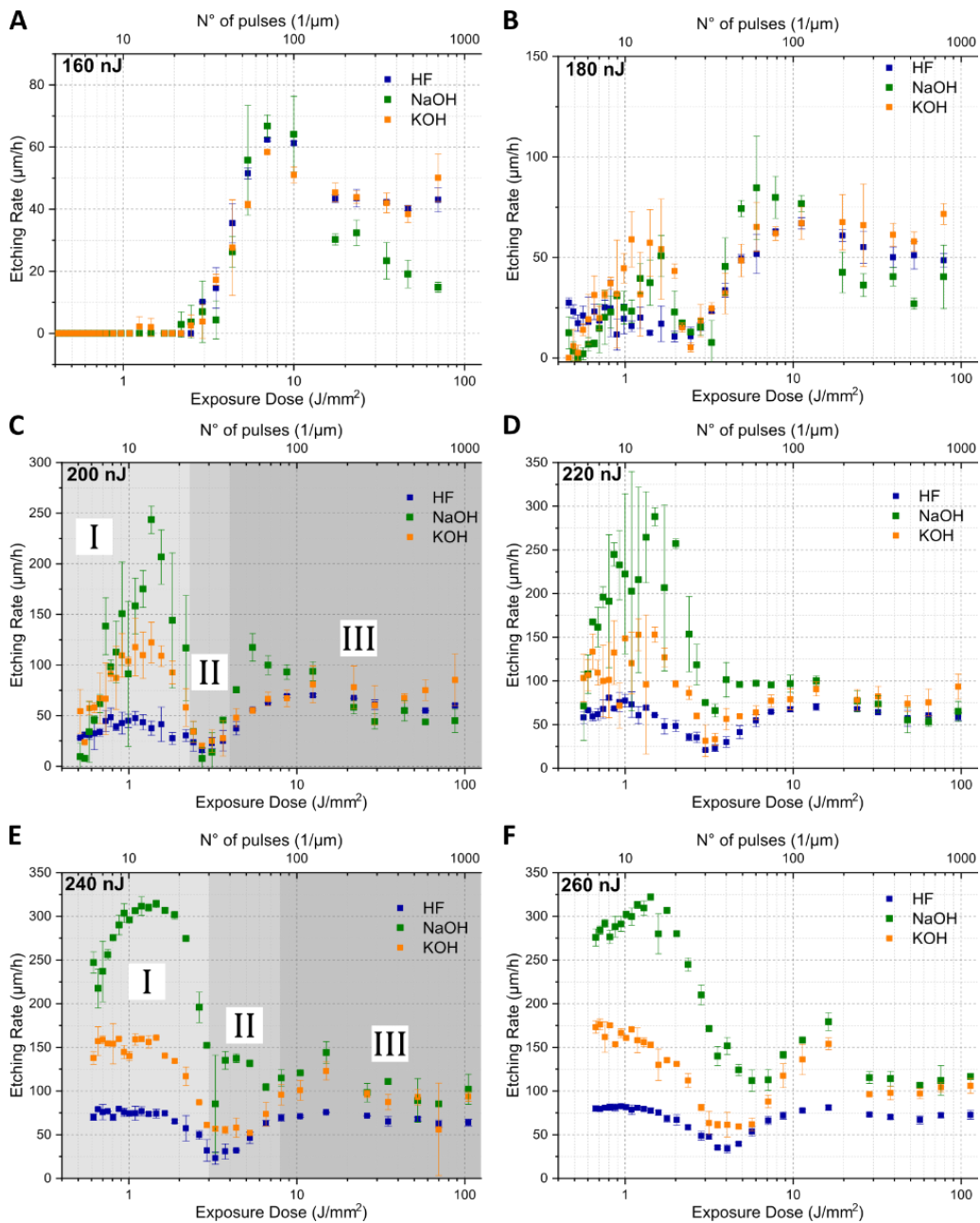

Fig. 1. Etching rate versus exposure doses for three different etchants: HF 2.5 % vol, KOH 45 % wt, and NaOH 5 % wt. Six different pulse energy levels are explored: 160 nJ (A), 180 nJ (B), 200 nJ (C), 220 nJ (D), 240 nJ (E), and 260 nJ (F). In each graph, the squares represent the mean value and the bars the standard deviation for the two measurements performed.

Each etchant naturally etches pristine fused silica, albeit at different rates. For similar etching conditions as in here, it is typically about ~ 3 µm/h for HF [8], it is about three times less for KOH (~ 0.9 µm/h) [17] and only ~ 0.5 µm/h for NaOH. To explore the effects of etching in terms of surface quality, a comparative study is performed and is provided in the Supplementary materials of this manuscript. In summary, HF leads to the least rough surfaces (Ra ~ 80 nm), while NaOH behaves similarly to KOH and results in a roughness around 50 % worse than the HF one (Ra ~ 120 nm). Finally, in Fig. 2C the performance of the most common subtractive glass micro-manufacturing methods are compared, taking as reference the highest aspect ratio values reported in [11,13,18–26]. The presented version of femtosecond laser assisted chemical etching with NaOH clearly stands out in terms of aspect ratio and unlike all the other techniques mentioned, enables the manufacturing of arbitrary 3D geometries, and this, independently from the etchant chosen.

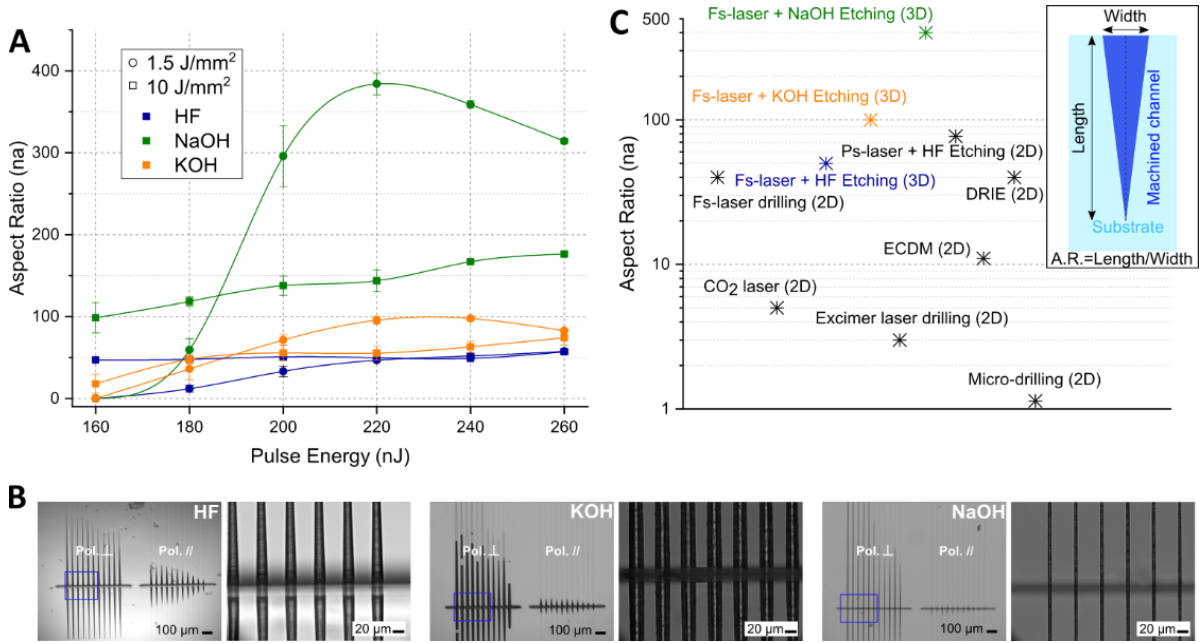

Fig. 2. (A) Aspect ratio of the etched lines at two characteristic exposure doses, for the three etchants and six different pulse energies. (B) Visual comparison of the etching progress for different laser parameters depending on the etchant used after 18 hours. (C) Comparison of the aspect ratio obtainable with various subtractive glass micromachining techniques. In the inset, a schematic of the aspect ratio definition used.

## 2.3 Effect of polarization

The polarization has a direct effect on the orientation of the nanostructures [9] and is known to affect the etching rate both for HF [10] and KOH [11]. Fig. 3 shows the etching polarization contrast, defined as the difference between etching rates for the two orthogonal linear polarization directions, defined as parallel and perpendicular to the writing direction. Like the other etchants, NaOH turns out to be also polarization sensitive. Interestingly, the polarization dependence drops to near zero (or even inverted) for doses corresponding to the low-etching-rate valley (~ 3-4 J/mm$^2$, see Fig. 1), where the material removal rate drops to a few tens of micron per hour. At very low doses, the difference between the two polarization states is the most pronounced and culminates at 300 µm/h for NaOH, for pulses of 240 nJ and doses of ~ 1 J/mm$^2$, value at which the etching rate for parallel polarization is near zero.

The etching rate contrast is actually the highest at very low doses despite the fact that only a handful of pulses do overlap, not allowing for a clear formation of self-organized nanogratings as visible through structural changes (see [15,16] and further in this manuscript). An important consequence of this observation is that there is an inherent anisotropy in the laser-affected zones *already* from the very first pulses, and well before nanogratings are formed.

The large polarization contrast has a direct implication on the process implementation as it requires ensuring that the polarization has consistently the same orientation with respect to the writing direction, which can for instance be achieved by mounting a half-wave plate in the beam path and rotating it to follow an arbitrary trajectory. A similar study is performed on the effect of the writing direction and is reported in the *Supplementary materials*. In particular, the writing direction differential remains between +/- 25 µm/h for HF, but reaches up to 100 µm/h for KOH and 150 µm/h for NaOH. This effect decreases for increasing deposited energy and might be due to the presence of a pulse-front-tilt in the laser beam.

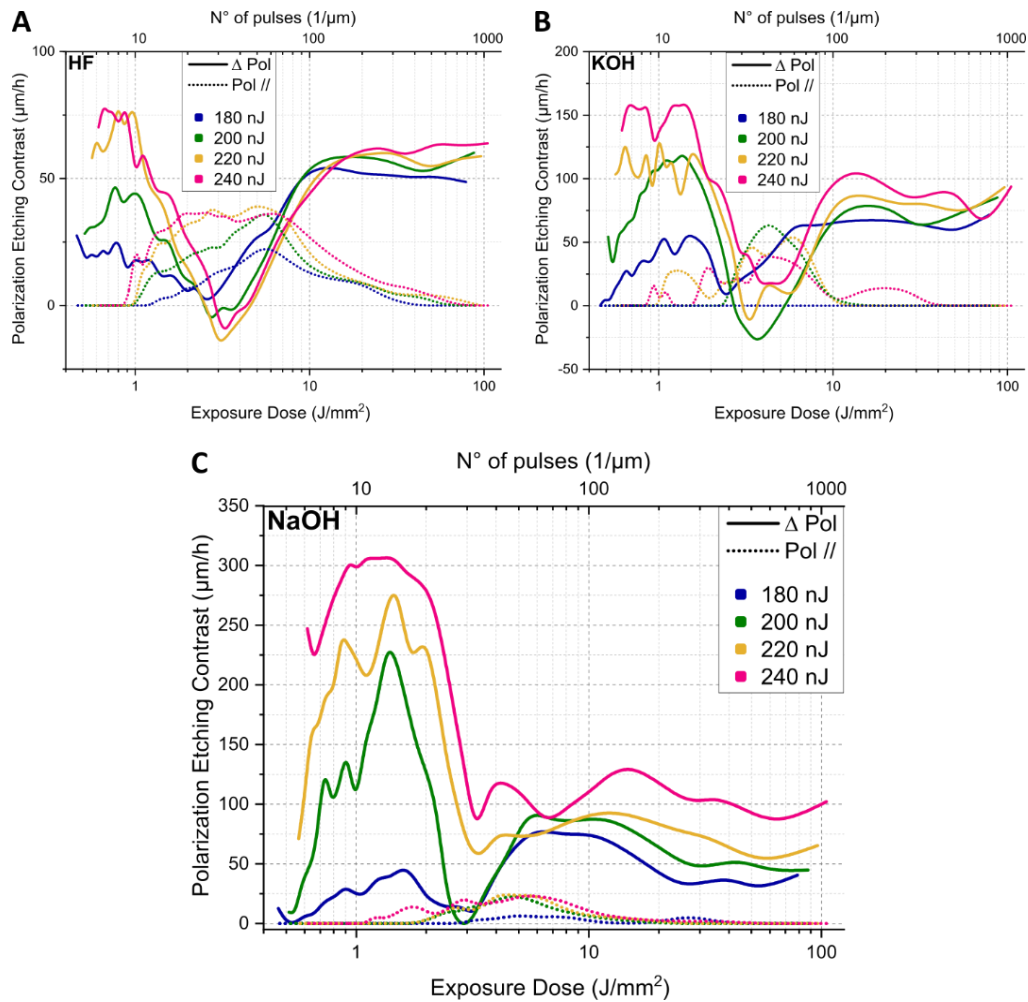

Fig. 3. Etching contrast between perpendicular and parallel laser polarization (defined with respect to the writing direction) versus exposure doses for different pulse energies and for the three etchants: (A) HF, (B) KOH, and (C) NaOH. For visual purposes, the lines represent the average trends based on the many experimental points. The solid lines show the etching contrast, while the dotted ones give the etching rate for parallel polarization.

*2.4 Cross-section analysis of laser-written patterns*

To further understand the nature of the material changes under varying pulse exposure conditions and to correlate it with etching observations, one sample with the same set of parameters tested is finely polished and etched in HF for three minutes to reveal the cross-sectional view of the laser written lines.

Pictures of the specimen are taken using a scanning electron microscopy (Zeiss® SEM Gemini 450 operated at 5.0 kV and 100 pA) and assembled in Fig. 4 to investigate morphological correlations between exposure dose and pulse energies with the etched patterns. In Fig. 4, the red-highlighted values correspond to the largest etching rate region for each pulse energy, while the blue-highlighted values to the energy deposited that leads to the formation of nanogratings, whether it is partial or complete. Interestingly, the highest etching rate does not match with the formation of well-structured nanogratings for pulse energies equal to or above 200 nJ. The first etching rate peak (found around 1.5 J/mm$^2$) has no direct correlation with morphological traits of nanogratings or possible micro-cracks. Interestingly, at very few overlapping pulses, the connection between porous zones is poor and does not form a continuous pattern as highlighted elsewhere [27,28]. Yet, the etching rate is among the highest observed suggesting an etching mechanism unrelated to the completeness of porous structures, interconnected or not.

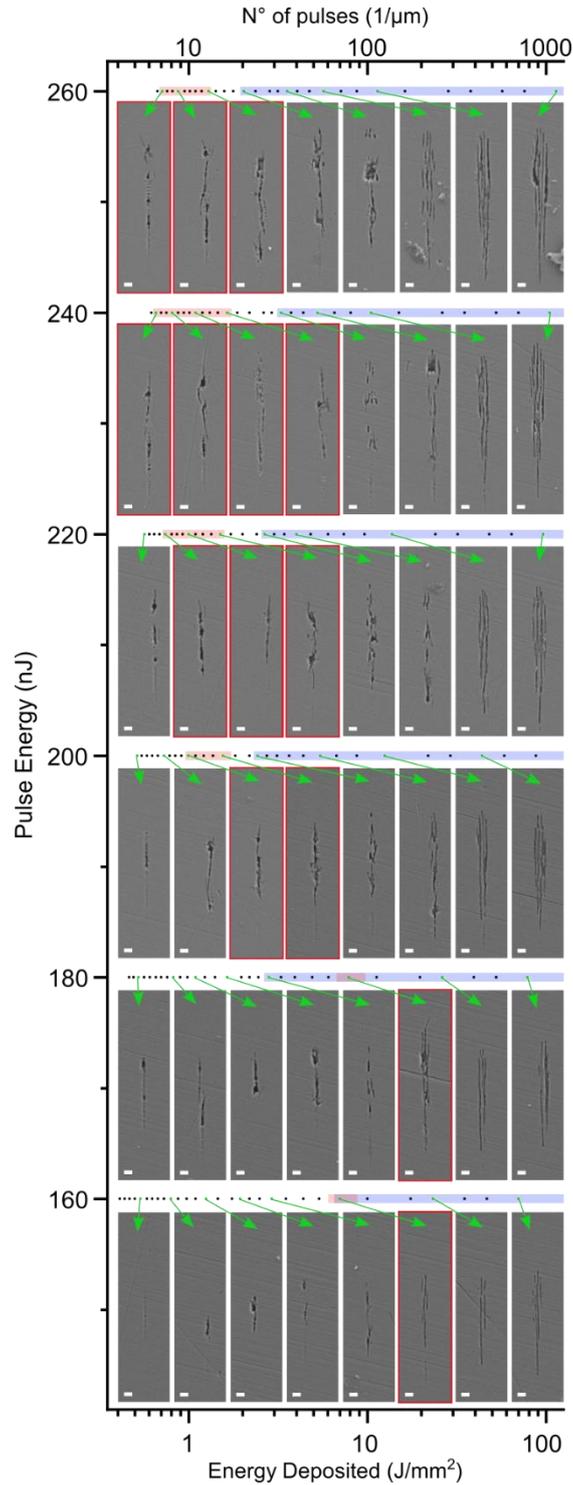

Fig. 4. Scanning electron microscope pictures of the cross-section of laser written lines for different pulse energy and energy deposited values. The squares represent the energy levels tested, with the green ones corresponding to the indicated images. The parameters highlighted in blue led to the formation of clearly visible nanogratings. The ones highlighted in red correspond to the largest etching rate. The scale bar is 1 µm.

*2.5 Effect of annealing on etching rates*

Non-ablative femtosecond laser interaction with fused silica leads to permanent defects in the glass-matrix when the material comes at rest. At a molecular level, oxygen deficiency centers (ODCs), non-bridging-oxygen-hole-centers (NBOHC), color centers (E' centers) and others[29] form and are considered precursors for the formation of free molecular oxygen trapped in nano-pores [30–32].

On one hand, nanogratings (forming 'microscopic *morphological* defects') resist to extreme annealing temperatures (above 1000°C) [33,34] (even though they undergo a slight degradation above 400°C [35]), while other localized laser-induced structural changes such as laser-induced densification, vanish around 900°C [36,37]. On the other hand, defects at the atomic glass matrix level that require much less energy to form, can be suppressed at 300°C [36] and be fully annealed at 500°C[30].

We take advantage of these observations to differentiate between morphological/structural versus glass-matrix defects as potential contributors for the observed accelerated etching, in particular in the low-dose exposure regimes. In practice, three specimens with a set of lines exploring different laser parameters were prepared as in the previous experiment, but considering only two pulse energies (200 and 240 nJ) and maintaining a polarization state perpendicular to the writing direction for optimal etching. The specimens were annealed at 300°C for 10 h, with a 1 °C/minute heating/cooling rate constant to limit the creation of defects during cooling [38]. Substrates were then etched for two hours in different etchants and the etching rate measured according to the same principle described in the previous paragraphs. Polishing was then performed to remove the portion of material containing etched lines. This procedure is repeated four times in order to test the effect of annealing at 300, 500, 700, and 900°C on the etching rate. The results are shown in Fig. 5.

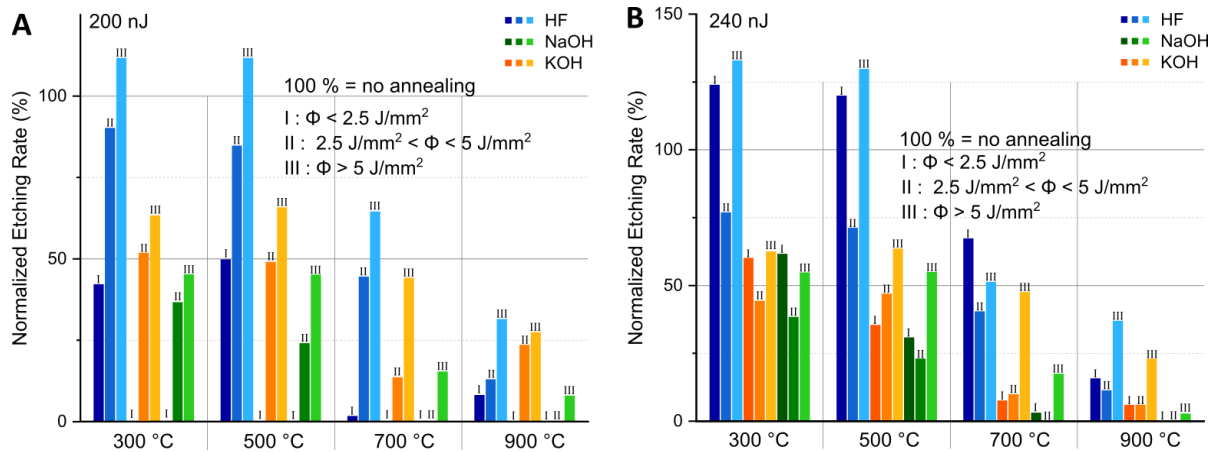

Fig. 5. Comparison of etching rates versus energy deposited before and after annealing for a fixed pulse energy of 200 nJ (A) and 240 nJ (B), for the three etchants (HF, KOH, and NaOH), and for the three distinct zones of interest defined in Fig. 1. The etching rate is normalized with respect to the etching rate *before annealing*.

Let us first consider the effect of the annealing on the HF etching rate (Fig. 5A-B, blue bars). The annealing up to 500°C – reported to quench defects [30,36] – reduces the etching rate only in the case of lower exposure doses and pulse energies (region I for 200 nJ), but does not decrease nor the higher-energy deposited peak nor the etching rate for a pulse energy of 240 nJ.

In fact, in both cases, we assist to a fictitious increase of the rate (only in region III for 200 nJ while in both region I and III for 240 nJ), which is due to a capillarity-driven degradation in etching selectivity of HF[11] and the decreased etching time from 4 h to 2 h for the measurements done before and after annealing, respectively. Such discrepancy among 200 and 240 nJ can be attributed to the creation of a larger amount of defects when higher pulse energy (i.e. field strength) is used, which can also explain the sharp decrease of noise in the etching rate measurements of Fig. 1 for increasing pulse energy and is in line with the stochastic essence of non-linear absorption phenomena. In other words, we assume that the decrease in defects' density, after annealing at 300°C and 500°C for 240 nJ of pulse energy, increases the mean distance between neighboring defects to a value that impacts minimally the HF.

The annealing temperature is further increased to values at which possible localized densification effects on the silica matrix are relaxed into larger and more stable ring structures (i.e. 700 and 900 °C) [34–37]. At this point, the HF etching performance degrades down to 50 and 30 % of the original etching speed before annealing. Such behavior hints that the HF etching mechanism is driven mainly by porosity and associated localized densification of the glass structure, at least in regions II and III. Indeed, structural changes locally cause a sharp increase in active surface and densification, suggested by the presence of small member rings, which augments the chemical reactivity of the Si-O bonds [39–41].

KOH (orange bars in Fig. 5) and NaOH (green bars in Fig. 5) show radically different behavior. At pulse energies of 200 nJ, the first annealing at 300°C almost completely cancels the low-energy deposited etching rate peak, while it halves it in the case of 240 nJ. As before, we interpret the different behaviors between pulse energies as due to the initial density of defects, larger for higher pulse energy, and assume that the annealing is not enough to cancel out all the defects for 240 nJ. Although to a lower extent, the etching peak corresponding to the highest energy doses (~ 10 J/mm$^2$) starts decreasing already at 300°C. When the annealing temperature is further increased, the decrease in etching rate continues smoothly for KOH and sharply for NaOH. Those trends indicate that the etching mechanism of KOH and NaOH is different from the one of HF and driven mainly by the presence of defects in the matrix and only partially by the presence of porosity or associated densification effects. In this context, NaOH tends to be the most selective and the most affected by the laser polarization state, the writing direction (see Fig. S3), or the annealing conditions. Finally, it should be noticed that even if mainly defects-driven the etching rate of NaOH and KOH is highly dependent on the laser polarization (as shown in Fig. 3B-C), which indicates that a self-organized structural configuration of the defects do exist at the very first stage of the laser interaction, and before the occurrence of pores and nanogratings forming a pathway for the etchant to progress.

We interpret these observations as follows. For low exposure dose (region I in Fig. 6), the etching enhancement is due to the presence of defects localized within the laser-affected-zone. With the pulse duration considered here, it is known that defects mostly consist of NBOHCs and E' centers (see in particular [42]). Both the cross-sectional analysis (Fig. 4) and the annealing study (especially for 200 nJ of pulse energy) supports this hypothesis. The lower (or none in case of HF) decrease in etching rate after 300°C and 500°C annealing of modifications with 240 nJ of pulse energy is linked to the density of defects created versus the amount canceled by the annealing. We assume that the mean-free(of defects)-path is reduced to a value that starts to influence the etching rate of KOH and NaOH, but not the one of HF, due to their difference in etching rate of pristine silica (see vectors of Fig. 6).

Further increasing the exposure dose (region II in Fig. 6) leads to a sharp decrease in etching rate which we attribute to the transition from defects to pores. However, the reason behind the dip in etching rate for region II remains at this stage unknown. Finally, in the high exposure dose (region III in Fig. 6) the nano-pores start coalescing and forming coherent networks with a preferential orientation and ultimately the so-called nanogratings. In this fashion, an accelerated etching is recovered, but this time is driven mainly by the inherent porosity of the material. Due to the difference in etching pristine material, in the case of hydroxide-based etchants, this second etching mechanism is significantly less favorable than the one driven by oriented defects, while it results in a similar enhancement for HF-based etchant.

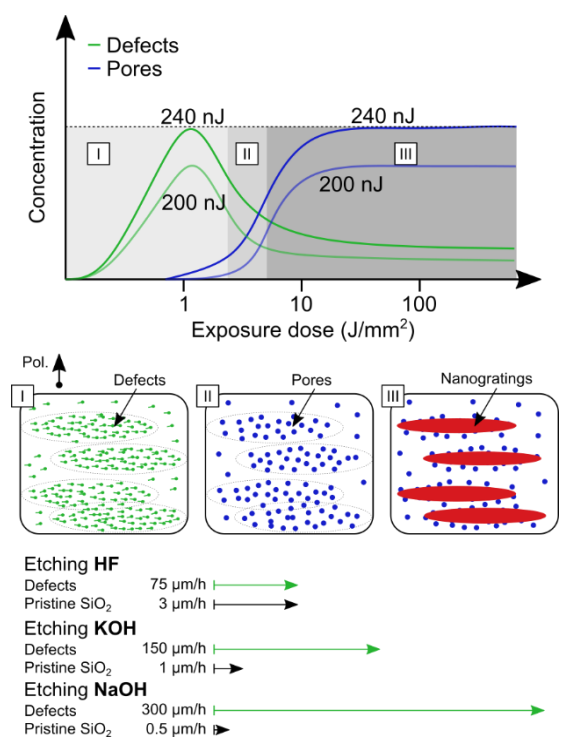

Fig. 6. Schematic interpretation of the etching mechanism in the region of interest defined above. Top to bottom: the graph shows a qualitative representation of the variation in defects' and pores' quantity depending on exposure dose and pulse energy; the three panels provide a visualization of the modifications induced in each regime with a fixed vertical polarization; and the colored vectors display the etching rate of dense of defects versus pristine $SiO_2$ for HF, KOH, and NaOH.

## 3. Summary and conclusion

Let us summarize the main findings of this study:
- For the three etchants investigated, a similar phenomenological behavior is observed: a first regime in the low-exposure doses corresponding to around ten overlapping pulses (~ 1.5 J/mm$^2$), followed by a second regime defined with a local minimum in etching rate and then a third regime, at higher dose, where an etching maximum is recovered.

- For the three-etchant investigated, the local etching maximum corresponding to the low exposure doses (~ 1.5 J/mm$^2$) with few pulses overlap, the so-called nanogratings are not yet distinctly observed. This first exposure regime at low doses is observed only above a given pulse energy threshold (here at ~ 180 nJ), higher than the one for observing modifications. Despite the absence of a visible self-organized pattern after etching, depending on the linear polarization orientation a strongly anisotropic etching behavior is observed. It is particularly pronounced in the case of NaOH, where an anisotropic ratio of ~ 1200 to 1 µm is observed after only 4 hours.
- HF acid-based etchant displays a different behavior than hydroxides-based etchants. While KOH and NaOH highest etching rates are observed for the low-exposure dose regime, for HF similar local maxima in etching rate are found in both the low- and high-exposure dose. Annealing below 500°C has a dramatic effect on the low-exposure dose etching behavior, eventually completely canceling the etching enhancement for NaOH and KOH at low pulse energy, while it has limited effects on the HF etching behavior.
- Sodium hydroxide (NaOH) as an etchant for femtosecond laser-assisted 3D micro-manufacturing shows superior performances compared to known alternative methods, based on HF and KOH solutions. It reaches an unprecedented etching rate of 300 µm/h, twice more than KOH and near four times more than HF. As the natural etching rate of silica by NaOH is very low, the contrast between exposed and unexposed etching rate yields an extreme aspect ratio approaching 1 to 400, for tunnel-like patterns.

In conclusion, from a technological point of view, the observation of an optimal etching rate surpassing the previously reported ones and this, at very low doses, offers a boost in achievable laser-exposure writing speeds larger than one order of magnitude. Moreover, while maintaining an optimal etching rate and selectivity, the aspect ratio obtainable is significantly increased, which is of practical importance for fabricating slender and denser structures.

**Funding sources and acknowledgments**

In this work, EC designed and carried out the experiments reported here. SP performed the initial preparatory experiments. YB and EC discussed and interpreted the results. YB conceived the research. The article text was mainly written by EC and YB. All authors contributed to revisions. The Galatea Laboratory acknowledges the sponsoring of Richemont International.

**Disclosures**

The authors declare no conflicts of interest.

**Data availability**

All data is available in the main text or in the Supplementary Materials.